\documentclass[%
 reprint,
 amsmath,amssymb,
 aps,
]{revtex4-1}

\usepackage{graphicx}
\usepackage{dcolumn}
\usepackage{bm}

\newcommand{\bra}{\langle}
\newcommand{\ket}{\rangle}
\newcommand{\plus}{{\!+\!}}

\newcommand{\be}{\begin{equation}}
\newcommand{\ee}{\end{equation}}
\newcommand{\bd}{\begin{displaymath}}
\newcommand{\ed}{\end{displaymath}}

\newcommand{\dsply}{\displaystyle}

\def\vecX{\mbox{\boldmath $X$}}

\def\vec0{\mbox{\boldmath $0$}}

\begin{document}

\preprint{APS/123-QED}
\title{
Solvable model of  a phase oscillator network  on a circle with infinite-range 
Mexican-hat-type interaction}
\author{Tatsuya Uezu$^1$},\email{uezu@ki-rin.phys.nara-wu.ac.jp} 
\author{Tomoyuki Kimoto$^2$} 
 \author{Masato Okada$^3$}
 \affiliation{
$^1$Graduate School of Humanities  and Sciences,
Nara Women's University, Nara 630-8506, Japan\\
$^2$Oita National College of Technology, Oita 870-0152, Japan\\
$^3$Graduate School of Frontier Sciences, The University of Tokyo, Kashiwa, Chiba 277-8561, Japan
}
\date{\today}
\begin{abstract}
We describe a solvable model of a phase oscillator network on a circle with  
 infinite-range  Mexican-hat-type  interaction.  
We derive self-consistent equations  
   of the order parameters  
 and obtain three non-trivial solutions characterized by the rotation number.  We also derive
 relevant characteristics such as
 the location-dependent distributions of the 
resultant frequencies of desynchronized oscillators.
Simulation  results closely agree with the theoretical ones.

\begin{description}
\item[PACS numbers]
05.45.Xt \ 
05.45.-a \ 
05.20.-y \ 
87.10.Rt \ 
\end{description}
\end{abstract}

\maketitle

It is ubiquitously observed in nature that
a system composed of many active elements exhibits collective behavior as a whole.
A typical example is the synchronization of populations of oscillators, e.g., 
 simultaneous emission of light by fireflies,
the rhythm of the  heart  composed of a population of  cardiac muscle cells, 
 and circadian rhythms \cite {saunders,Cloudsley-Thompson}.\par
Pioneering studies on such behavior were done 
by Winfree \cite{winfree} and Kuramoto \cite{kuramoto-book}.
In particular, Kuramoto regarded synchronization  as
a phase transition and described 
a prototype model of the phase transition in non-equilibrium systems.
 The model, today   called the ``Kuramoto model,'' is a coupled oscillator system
 in which an oscillator interacts with  all other oscillators  with the same strength.
Each oscillator has its own natural frequency, but its amplitude is constant and
the state  variable is its phase. 
In general, when nonlinear dynamical systems with  stable limit cycle oscillators 
 are weakly coupled, the whole system can be
described by the phases of the oscillators, and the dynamical equation
 is reduced to the evolution equation for phases \cite{kuramoto-book}.
  The Kuramoto model was used to 
 analytically prove for the first time that, 
as the interaction strength  increases from zero, 
 a phase transition  occurs,  
from a desynchronized state in which each oscillator
independently oscillates with its own frequency 
 to a synchronized state 
in which a large number of oscillators oscillate with the same frequency \cite{kuramoto-1}.
 Since  Kuramoto's analysis of  globally coupled oscillators, 
 oscillator networks  with short-range and with intermediate-range interactions
 have been studied \cite{sakaguchi.etal}. Oscillators with global and
 random interactions \cite{daido} 
and with sparse and random interactions \cite{hatchett.uezu-1} 
have also been studied.  Furthermore, the stability of the solutions with the Kuramoto model
has been studied \cite{Mirollo.Strogatz,Chiba}. 
A review of the Kuramoto model and its extensions,
 such as inclusion of a noise term, is available elsewhere \cite{rev.mod.phys}.
There have also been extensive studies on the statistical and dynamical  properties 
of the mean-field XY model (HMF XY model) of conservative dynamical systems corresponding 
 to oscillaor network models of  
dissipative dynamical systems \cite{antoni.ruffo,Campa.etal}.\par
Although there have been many studies on oscillator networks, 
no solvable model  defined in a finite-dimensional space
has yet been  introduced. 
It is greatly  difficult  
to study systems with short- and intermediate-range interactions
analytically, so we cannot help relying on numerical simulation to study such systems. 
To  further advance the study of the synchronization--desynchronization
transition of active elements, it is quite desirable to introduce a solvable 
model that extends the Kuramoto model.\par
In this paper, we describe a phase oscillator network on a circle,
and, to make analytical treatment possible, 
we assume  infinite-range  interaction and that the strength and
 sign of the interaction between two oscillators 
 depend on the spatial distance between them. 
We specifically adopt the Mexican-hat-type interaction, which 
 was introduced to model the creation of feature extraction cells
in  neuroscience and expresses the  properties that 
a firing  cell  excites nearby cells and inhibits distant cells \cite{Hubel.Wiesel}.
For an XY model on a circle with this interaction,
it was found that
 there exists a peculiar solution, the pendulum solution, in which
the phases of the XY spins do not rotate but oscillate
 as the locations change on the circle \cite{kimoto.etal}. 
In the phase oscillator network,
we show that the self-consistent equations (SCEs) of the 
order parameters for stationary states and the relevant quantities 
can be exactly derived theoretically, and that 
there exists a pendulum solution as in the XY model.\par
Now we explain the phase oscillator network.
Let  $\phi_i$  and $\theta_i$ be  the phase and location on the circle
 of the  $i$-th oscillator.
We regard the $i$-th oscillator as a two-dimensional vector, $\vecX_i=(\cos \phi_i, \sin \phi_i)$.
We assume that oscillators are located uniformly on the circle;  
that is,  $\theta_i= i 2 \pi/N \ ( i=0, \cdots, N-1)$.
The evolution equation for the $i$-th phase is
\begin{eqnarray}
\frac{d}{dt} \phi_i &=& \omega _i + \frac{1}{N} \sum_j J_{ij}\sin(\phi_j
- \phi_i),
\label{eq:evolution}
\end{eqnarray}
which is derived under the rather general situation described above.
 Here,  $\omega_i$ is the natural frequency and is drawn from the
probability density $g(\omega)$, which is assumed to be
one-humped at $\omega = \omega_0$ and symmetric with respect to $\omega_0$.
In our numerical simulations, we used a Gaussian distribution $g(\omega)$ with a mean of zero and 
 a standard deviation $\sigma$.\par  
Let us explain the interaction we use in this paper in detail. 
We impose  translational symmetry  on $J_{ij}$, i.e.,  
 $J_{ij}$ takes the form $J_{ij}=J(\theta_i -\theta_j)$.
Furthermore, we assume $J_{ij}=J_{ji}$. 
Thus,  $J(\theta)$ is  an  even function of $\theta$.
Therefore, the Fourier expansion of $J(\theta)$ is given by
\begin{eqnarray}
J(\theta) &=& J_0+J_1\cos(\theta) + J_2\cos(2 \theta) + \cdots .
\end{eqnarray}
In this study, we treat the case in which  only $J_0$ and $J_1$ are non-zero, so 
 the  interaction is
\begin{eqnarray}
J_{ij} &=& J_0+J_1 \cos(\theta_i-\theta_j),
\end{eqnarray}
which has the properties expressed by  the Mexican-hat-type interaction described above.

The order parameters are defined as
\begin{eqnarray*}
&& R e^{i \Theta}   =  \frac{1}{N}\sum_j   e^{i\phi_j},\\
&& R_c e^{i \Theta_c} = \frac{1}{N} \sum_j \cos \theta_j  e^{i\phi_j},\
R_s e^{i \Theta_s} = \frac{1}{N} \sum_j \sin \theta_j  e^{i\phi_j}. 
\end{eqnarray*}
Using these order parameters, we rewrite the evolution equation (\ref{eq:evolution}) as
\begin{eqnarray}
\frac{d}{dt} \phi_i &=& \omega _i + 
 J_0    R \sin( \Theta - \phi_i) \nonumber\\
&&\hspace{-1.5cm} +  J_1 [ R_c \cos \theta_i \sin( \Theta_c -\phi_i)
+ R_s \sin \theta_i \sin (\Theta_s - \phi_i) ].
\label{dif.eq}
\end{eqnarray}
Now we derive the SCEs.  Without loss of generality, we assume $\omega_0=0$.
Since we study  stationary states, let us assume that amplitudes and phases tend to
 constant values as $t$ tends to infinity.
  We further rewrite eq. (\ref{dif.eq}) as
\begin{eqnarray}
\frac{d}{dt} \phi_j &=& \omega _j -
 A_j  \sin(\phi_j- \alpha_j).
\label{dif.eq2}
\end{eqnarray}
The following relation is derived from a comparison of eqs. (\ref{dif.eq}) and (\ref{dif.eq2}):
\begin{eqnarray}
A_j e^{i \alpha_j} &= & 
 J_0    R e^{i \Theta }
 +  J_1 [ R_c  \cos \theta_j e^{i \Theta_c }
+ R_s  \sin \theta_j e^{i \Theta_s }].
\label{comp.op}
\end{eqnarray} 
Hereafter, we use $\theta$ to identify each oscillator, so $A_\theta$ is expressed as
\begin{eqnarray}
&& A_{\theta}^2 = (J_0 R)^2
  + J_1 ^2 \{  (R_{c}\cos \theta)^2 +( R_{s}\sin \theta )^2 \nonumber \\
&&  + 2R_{c} R_{s} \cos(\tilde{\Theta}_{c}- \tilde{\Theta}_{s})
 \sin \theta \cos \theta  \} \nonumber \\
&& + 2 J_0 J_1 R \{ R_{c } \cos \tilde{\Theta}_{c} \cos \theta +
 R_{s } \cos \tilde{\Theta}_{s} \sin \theta \}, 
\label{Xi}
\end{eqnarray}
where $\tilde{\Theta}_{c} \equiv \Theta_{c} - \Theta, \ \ 
  \tilde{\Theta}_{s} \equiv \Theta_{s} - \Theta.$
Defining $\psi_\theta \equiv \phi_\theta - \alpha_\theta$ transforms the evolution equation into
 \begin{eqnarray}
\frac{d}{dt} \psi_\theta &=& \omega _\theta - A_\theta  \sin \psi_\theta.
\label{dif.eq3}
\end{eqnarray}
From this equation, we can develop a theory by following Kuramoto's argument.
For the synchronized oscillators satisfying $|\omega_\theta| \le A_\theta$,
 we obtain the entrained phase $\psi^* _\theta$ 
and the number density of the synchronized oscillators with the value of phase $\psi$ at 
 location $\theta$,
 $n_s(\theta, \psi)$, as
\begin{eqnarray}
\hspace{-1cm}
\psi^* _\theta &=& {\rm Sin} ^{-1} (\frac{\omega_\theta}{ A_{\theta}}), \\
n_s(\theta, \psi) 
&=& g(A_{\theta} \sin \psi)A_{\theta} \cos \psi, 
  \ \  |\psi| \le \frac{\pi}{2},
\end{eqnarray}
where Sin$^{-1} x$ is the principal value and its range is 
$\dsply [ -\frac{\pi}{2}, \frac{\pi}{2}]$.
For the desynchronized oscillators satisfying 
$|\omega_\theta| > A_\theta$, we obtain the solution of 
differential equation (\ref{dif.eq3}) and the number density of the desynchronized 
oscillators with the value of phase $\psi$ at  location $\theta$, $n_{ds}(\theta, \psi)$, as
\begin{eqnarray}
\hspace{-1cm}
\psi_\theta &=& \tilde{\omega}_\theta t 
+ h(\tilde{\omega}_\theta t),\\
\tilde{\omega}_\theta  &= & 
\omega_\theta \sqrt{1-(\frac{A_\theta}{\omega_\theta})^2},\\
n_{ds}(\theta, \psi)
&=& \frac{1}{ \pi} \int_{A_{\theta}} ^{\infty} \ dx \
 x \ g(x) \frac{\sqrt{x^2-A_{\theta} ^2}}{x^2-A_{\theta} ^2 
\sin^2 \psi},
\label{both.nofds}
\end{eqnarray}
where $\tilde{\omega}_\theta$ is the resultant frequency and $h(t)$ is a
periodic function of $t$ with period $2\pi$.
Note that the entrained phases and the distribution of resultant
frequencies depend on the oscillator locations, in general.
 From eq. (\ref{both.nofds}), 
$n_{ds}(\theta, \psi + \pi) = n_{ds}(\theta, \psi)$ is derived, 
and  $ \int_0 ^{2\pi} n_{ds}(\theta, \psi) e^{i \psi} d \psi =0$ follows.
Thus, only the synchronized oscillators contribute to the order parameters:
\begin{eqnarray}
 R e^{i \Theta}  
& = &
\int_{-\pi}^{\pi}  d \psi   n_s(\psi)   e^{i \psi + i \alpha_{\theta}},
\label{ceq.r}\\
R_c e^{i \Theta_c} &=&
\int_{-\pi}^{\pi} d \psi \frac{1}{2 \pi}\int_{0} ^{2\pi}
 d \theta n_s(\theta, \psi)
\cos \theta   e^{i \psi + i \alpha_{\theta}},
\label{ceq.rc}\\
R_s e^{i \Theta_s} &=&
\int_{-\pi}^{\pi} d \psi \frac{1}{2 \pi}\int_{0} ^{2\pi} 
d \theta n_s(\theta, \psi)
\sin \theta   e^{i \psi + i \alpha_{\theta}},
\label{ceq.rs}
\end{eqnarray}
where $\dsply  n_s(\psi)=
\frac{1}{2 \pi}\int_{0} ^{2\pi} d \theta n_s(\theta, \psi)$.
Substituting the expression for $n_{\rm s}(\theta, \psi)$ into these equations,
and after some algebra, from the real parts of these equations, we obtain
\begin{eqnarray}
 R & = &
 J_0  R \bra 1 \ket +  J_1 ( R_c f \cos  \tilde{\Theta}_c  
+ R_s g \cos  \tilde{\Theta}_s  ),
\label{R} \\
R_c 
 &=&
 J_0  R f \cos \tilde{\Theta}_c 
+  J_1 \{ R_c  a 
+ R_s c \cos (\tilde{\Theta}_c - \tilde{\Theta}_s)    \},
\label{Rc} \\
R_s  &=&
 J_0  R g \cos \tilde{\Theta}_s
 +  J_1 \{ R_c c  \cos (\tilde{\Theta_c} - \tilde{\Theta}_s) 
  + R_s  b \}, 
\label{Rs} \\
&&   a=\bra \cos ^2 \theta \ket,\  b=\bra \sin ^2 \theta \ket, 
 c=\bra \sin \theta \cos  \theta \ket, \nonumber \\
&&  f =\bra \cos \theta \ket, \ g =\bra \sin \theta \ket, \nonumber \\
&& \bra B \ket = \frac{1}{ \pi}
\int_0 ^{\pi/2}  d \psi 
\int_{0} ^{2\pi} d \theta 
g(A_{\theta} \sin \psi) \cos^2 \psi \ B. \nonumber
\end{eqnarray}
These are the SCEs for $R, R_c$, and  $R_s$.  Furthermore, we derive
the following auxiliary equations 
 from the imaginary parts of eqs. (\ref{ceq.r})-(\ref{ceq.rs}):
\begin{eqnarray}
 J_0  R f \sin \tilde{\Theta}_c 
+  J_1  R_s c \sin (\tilde{\Theta}_c - \tilde{\Theta}_s)   &=& 0  ,
\label{imRc} \\
 J_0  R g \sin \tilde{\Theta}_s
 -  J_1  R_c c  \sin (\tilde{\Theta_c} - \tilde{\Theta}_s)
  &=& 0.
\label{imRs}
\end{eqnarray}
 From these equations, the phases of
the order parameters are  completely determined.
 The detailed results will be reported elsewhere.
There are four solutions of the SCEs,  and they are classified on the basis of the
values of $R$ and $R_1=\sqrt{R_c^2+R_s^2}$ as
\begin{eqnarray*}
&& \mbox{ P: para magnetic solution, $(R, R_1)= (0, 0)$},\\
&& \mbox{ U: uniform solution, $(R, R_1)= (+, 0)$},\\
&& \mbox{ S: spinning solution, $(R, R_1)= (0, +)$}, \\
&& \mbox{ Pn: pendulum solution, $(R, R_1)= (+, +)$}.
\end{eqnarray*}\par
Now, let us consider the  physical meanings of these solutions.
To characterize the solutions further, we define the rotation number of a solution.
The rotation number is the number of rotations 
of synchronized oscillator  $\vecX_\theta ^*  =(\cos \phi_\theta ^*, \sin \phi_\theta ^*)$
 around the origin in  space  $\vecX$
 as  location $\theta$ changes by $2\pi$.
In the P solution, all oscillators
desynchronize, whereas in the other three solutions, 
an extensive number of oscillators synchronize and their directions are locked.
In the U solution,  $\phi_\theta ^*$ 
randomly takes a value in the interval $\dsply[- \frac{\pi}{2}+\Theta, \frac{\pi}{2}+\Theta]$
 irrespective of the location of the oscillator, so the rotation
 number is 0.
In the S solution, $\phi_\theta ^*$  linearly depends 
 on $\theta$, and the rotation number is 1.
In the Pn solution,  $\phi_\theta ^*$
  fluctuates and the rotation number is 0. See Figs. 1(a)-(c).\\
\begin{figure}[ht]

\begin{picture}(0,85)
\put(-145,0){\includegraphics[width=3.3cm,clip]{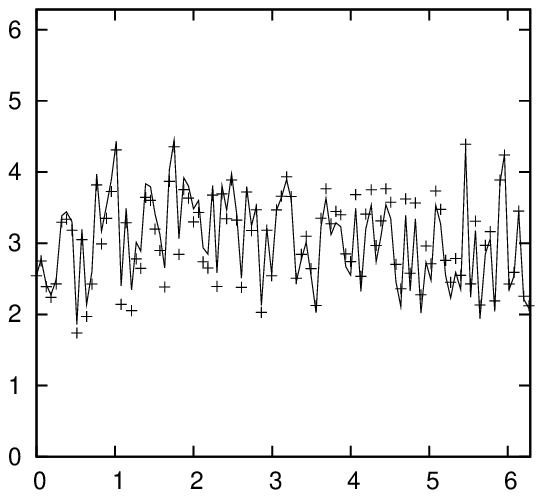}}
\put(-95,79){(a)}
\put(-150,40){$\phi^* _\theta$}
\put(-95,-5){$\theta$}
\put(-55,0){\includegraphics[width=3.3cm,clip]{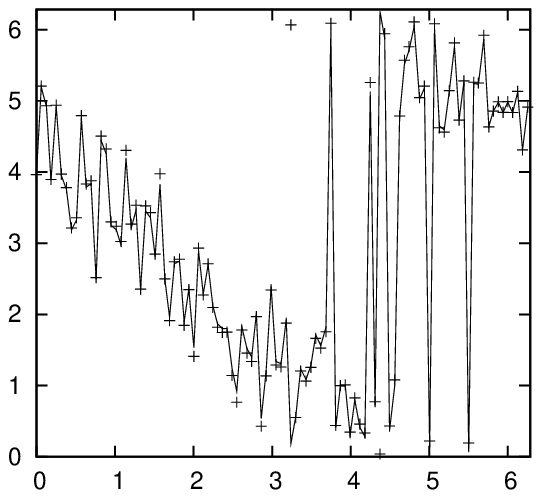}}
\put(-5,79){(b)}
\put(-55,40){$\phi^* _\theta$}
\put(-5,-5){$\theta$}
\put(35,0){\includegraphics[width=3.3cm,clip]{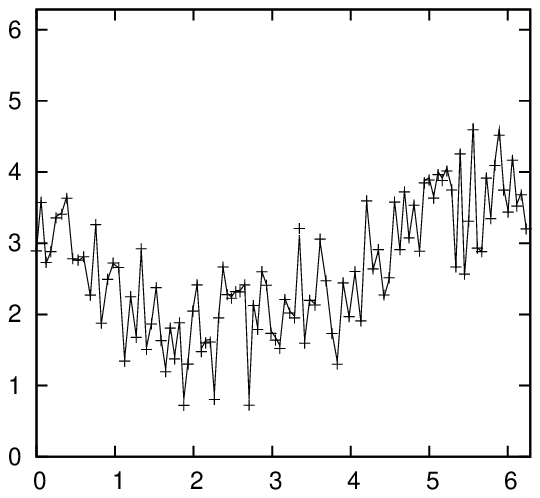}}
\put(85,79){(c)}
\put(35,40){$\phi^* _\theta$}
\put(85,-6){$\theta$}
\end{picture}
\caption{$\theta$ dependencies of entrained phases $\phi^* _\theta$.
 Line plots: theory; $\plus$: simulation ($N=10000, \sigma=0.2, J_0= 1.2 J_{0,c}$).
 Only 1$\%$ of entrained phases are depicted.
(a) U solution, $J_1/J_0=1.9$,
(b) S solution, $J_1/J_0=2.1$, (c) Pn solution, $J_1/J_0=2.1$.
}\label{fig.1}
\end{figure}
\begin{figure}[ht]
\begin{picture}(0,80)
\put(-65,90){(a)}
\put(-110,-10){\includegraphics[width=3.5cm,clip]{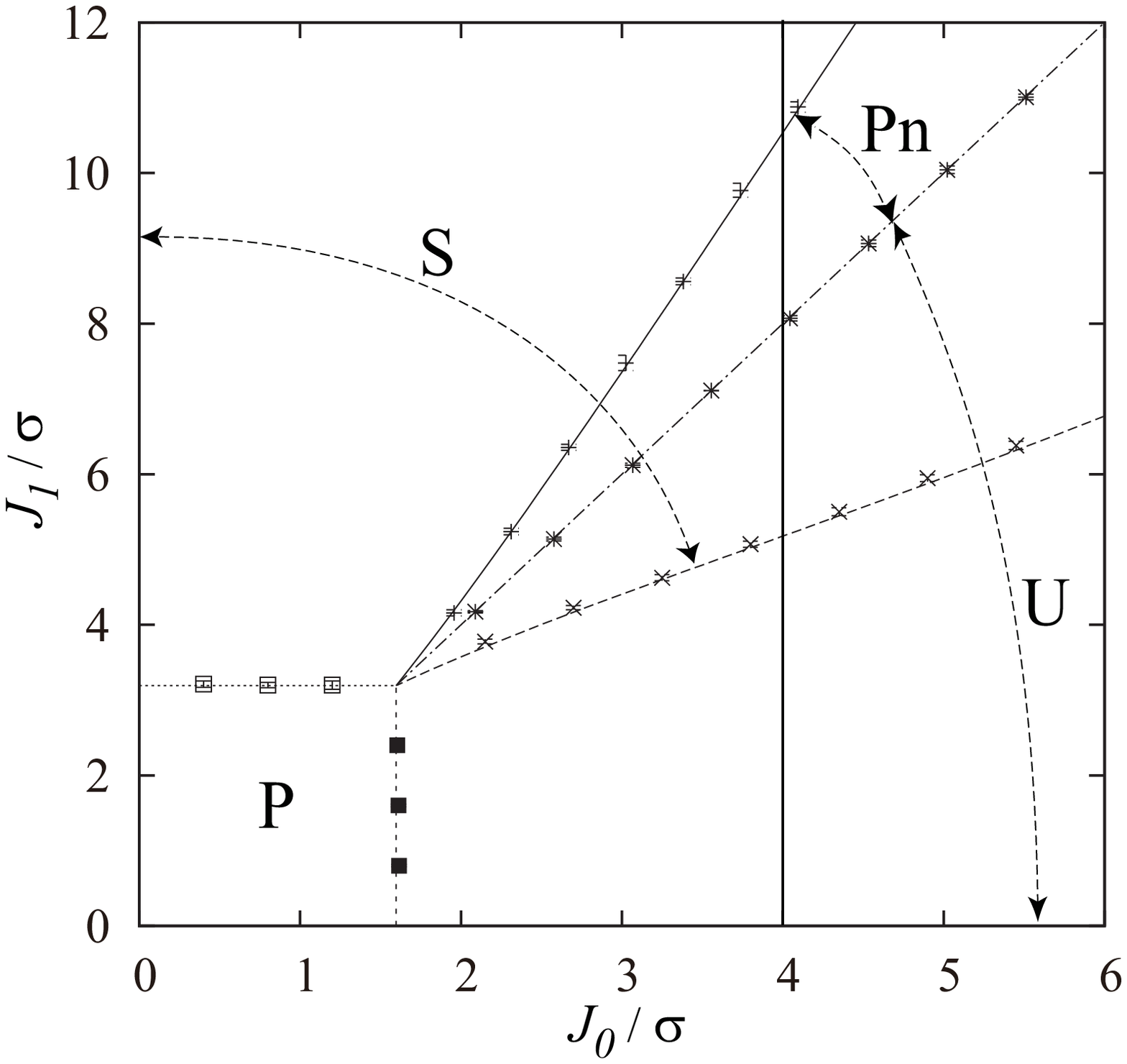}}
\put(55,90){(b)}
\put(10,-10){\includegraphics[width=3.5cm,clip]{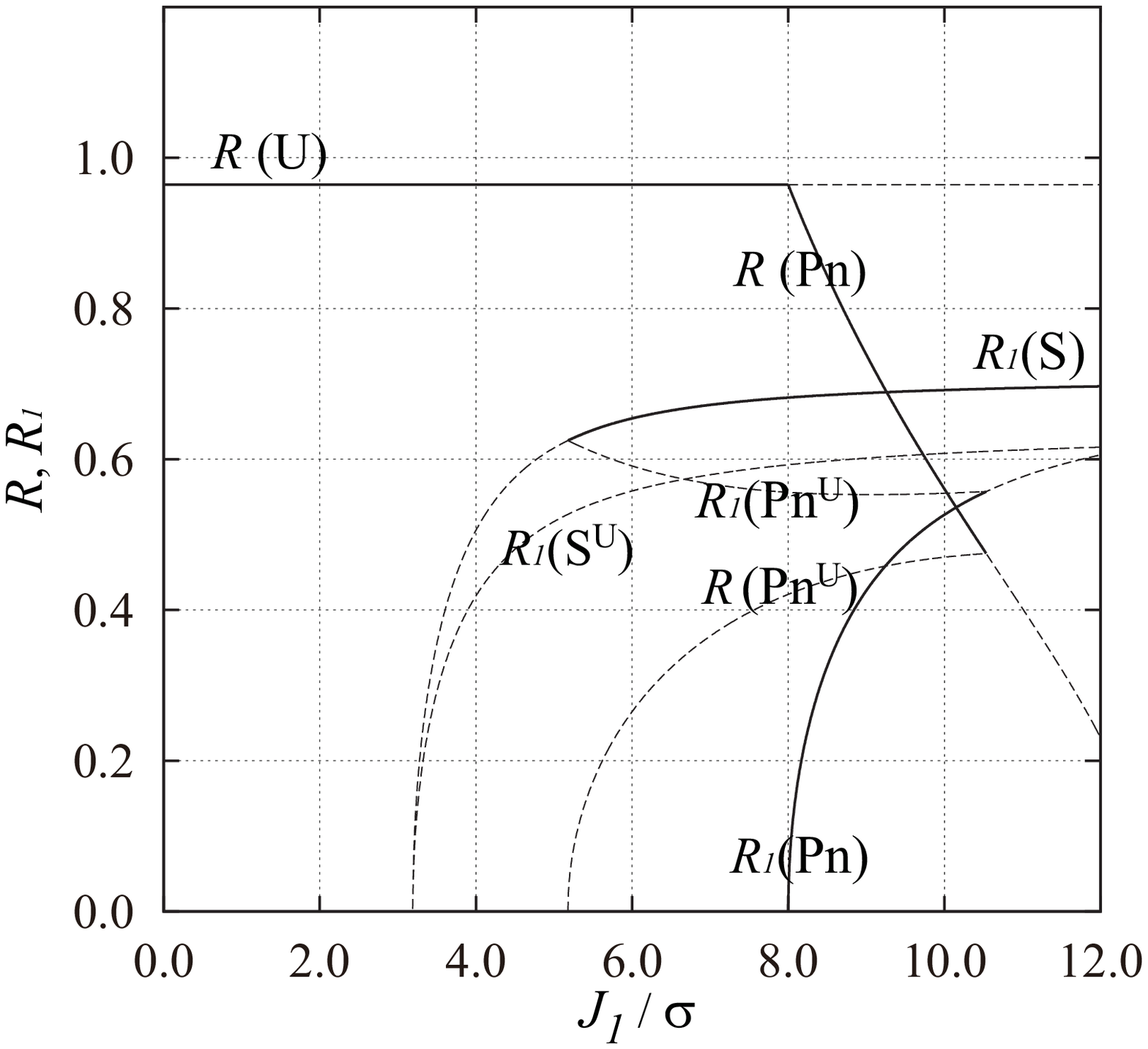}}
\end{picture}
\caption{(a) Phase diagram in  scaled parameter space. 
Plot points represent simulation results; curves represent theoretical results.
Vertical line represents parameters shown in  (b).
(b) $J_1/\sigma$ dependencies of order parameters.
 $ J_0/ \sigma=4$.
Solid curves represent stable solutions; dashed curves represent unstable solutions,
 which have superscript U, e.g., S$^{\rm U}$.
}\label{fig.2}\vspace{-0.5cm}
\end{figure}
We display the phase diagram in the scaled parameter space
$(J_0/\sigma, J_1/\sigma)$ in Fig. 2(a).\par
Let us examine  the bifurcation structures.
As shown in the phase diagram in Fig. 2(a), the S and U solutions  
and the S and Pn solutions can coexist.  We display the
 $J_1/\sigma$ dependencies of order parameters
 $R$ and   $R_1$ with $ J_0/ \sigma$ fixed to 4 in Fig. 2(b).
These results show that the unstable Pn solution determines the boundary
of the coexistent regions of the S and U solutions and  that of the S and Pn solutions.
Furthermore, we note  that the Pn solution continuously 
bifurcates from the U solution.  
Taking into account these observations, 
we derived the formulas for the  boundaries of 
 bistable regions  by using the unstable Pn solution 
and relevant stable solutions. 
In Fig. 2(a), the  theoretically obtained  boundaries are represented  by curves.
The theoretical results are in good agreement
  with the  simulation results.  \par
Now, let us examine the physical meanings of the phase transitions.
There are five boundaries in the phase diagram shown in Fig. 2(a).
The transition from the P to U phase takes place continuously at 
$J_0=J_{0,c} \equiv 2/(g(\omega_0) \pi)$, and this
 is the same transition as in the Kuramoto model.
The transition from the P to S phase takes place continuously at 
$ J_1=J_{1,c} \equiv 2 J_{0,c}$. 
In the P phase, the rotation number is not defined while it is 1 in the S phase.
The transition from the U to Pn phase takes place continuously at $J_1=2J_0$.
In this case, the rotation number in  both phases is 0.
However, as is shown in Figs. 5(a) and (c), 
 the directions of  two synchronized oscillators do not correlate in the
U phase but correlate weakly
in the Pn phase because  magnitude $J_1$ of the location-dependent
interaction is larger after the transition than before the transition.
At the two bistable region boundaries, 
the stable S  and unstable Pn solutions 
and  the stable Pn and unstable Pn solutions 
merge, and the stable S and  stable Pn solutions disappear.
 The unstable Pn solution  differs from the paired solution of the
stable Pn solution because the phases of the order parameters are different.
That is, these are a new type of instability that does not exist in the Kuramoto model.\\
\begin{figure}[ht]
\begin{picture}(0,60)
\put(-80,-5){\includegraphics[width=3.5cm,clip]{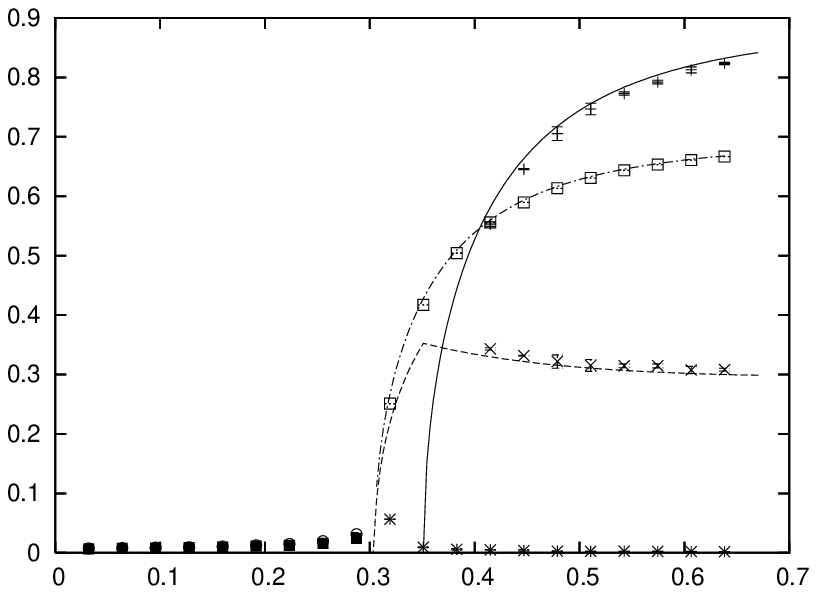}}
\put(-85,40){$R$}
\put(-85,30){$R_1$}
\put(-30,-12){$J_0$}
\end{picture}
\caption{
 $J_0$ dependencies of order parameters. $J_1=2.1J_0$.
Curves represent theory; symbols represent simulation results.   $N=20000, \sigma=0.2$. 
Averages were taken  for 20 samples.
Solid curve and +: $R$ of Pn solution; 
dashed curve and $\times$: $R_1$ of Pn solution; 
dashed dotted curve and square: $R_1$ of S solution.
Vertical lines are error bars. 
}
\label{fig.3}
\end{figure}
\begin{figure}[ht]
\begin{picture}(0,60)
\put(-58,65){(a)}
\put(-113,-5){\includegraphics[width=3.5cm,clip]{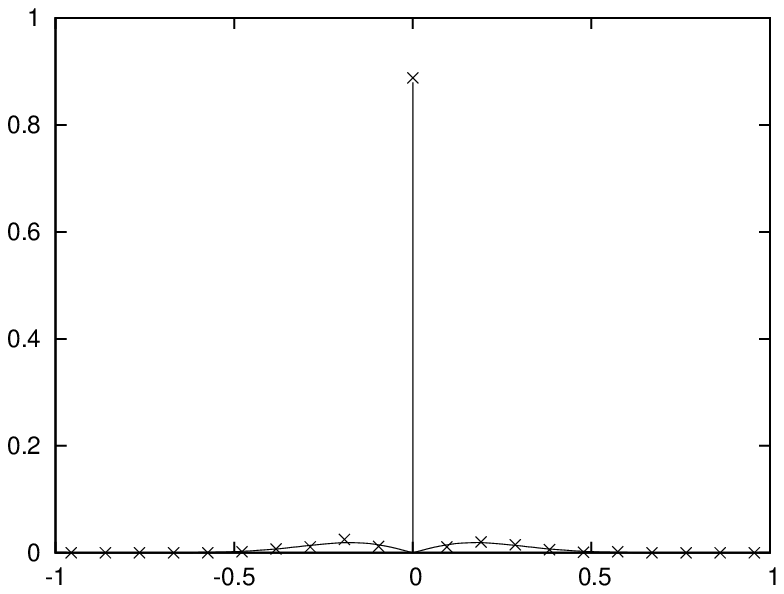}}
\put(-140,25){$G(\tilde{\omega}, \theta)$}
\put(-58,-10){$\tilde{\omega}$}
\put(8,-5){\includegraphics[width=3.5cm,clip]{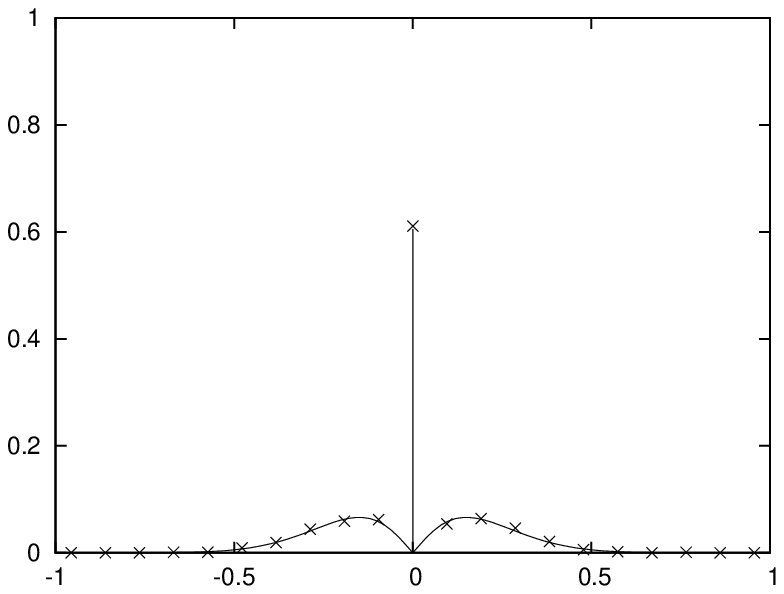}}
\put(58,65){(b)}
\put(62,-10){$\tilde{\omega}$}
\end{picture}
\caption{
Distribution $G(\tilde{\omega},\theta)$ 
of  resultant frequencies
 for Pn solution.
Curve represents theory; $\times$ represents simulation results ($N=100000, \sigma=0.2, J_1/J_0=2.1, J_0= 1.2 J_{0,c}$).
(a) $\theta=0.05 \times 2 \pi$, (b) $\theta=0.25 \times 2 \pi$.
}\label{fig.4}
\end{figure}
We show theoretical and numerical results for 
the $J_0$ dependencies of the order parameters 
in Fig. 3, those for the location-dependent 
resultant frequency distribution $G(\tilde{\omega}, \theta)$ for different $\theta$ 
for the Pn solution in Fig. 4,
and those for the $\theta$ dependencies
 of the entrained phases $\phi^* _\theta$ for the U, S, and Pn solutions in Figs. 1(a)-(c).
The agreement between the theoretical and numerical results is excellent.
 To investigate the  desynchronized oscillators, 
we constructed a Lorenz plot of  time series  $\sin(\phi_i(t))$ for the Pn solution (Fig. 5).
\begin{figure}[ht]
\begin{picture}(0,70)
\put(-50,-5){\includegraphics[width=3.5cm,clip]{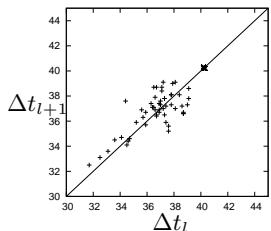}}
\put(-55,35){$\Delta t_{l+1}$}
\put(0,-10){$\Delta t_{l}$}
\end{picture}
\caption{Lorenz plot  for  desynchronized oscillator
for  Pn solution. 
$\times$: theory; $\plus$: simulation ($N=10000, \sigma=0.2, J_1/J_0=2.1, J_0= 1.2 J_{0,c}$).
}\label{fig.5}
\end{figure}
The Lorenz plot is defined as the mapping from  the difference $\Delta t_{l}=t_{l+1}-t_l$ 
to $\Delta t_{l+1}$, where 
 $t_l$ and $t_{l+1}$ are the successive times that satisfy $\cos((\phi_i(t_l))=1$
and $\cos(\phi_i (t_{l+1}))=1$, respectively.
As shown in Fig. 5, the simulation results are scattered in the Lorenz plot.
This indicates that the trajectory of a desynchronized oscillator
behaves  chaotically even though theoretically it is quasi-periodic.
However, in  most of our numerical results, e.g., for the resultant frequency distribution,
 the larger  the $N$, the better 
the agreement between the theoretical and numerical results. 
Our results suggest that 
the system behaves quasi-periodically in the limit of $N$ infinity. \par
In summary, 
we have extended the Kuramoto model,  which
is a prototype model of the synchronization--desynchronization phase
 transition
in non-equilibrium systems, and 
have proposed  a solvable model of a  phase oscillator network on 
 a circle with  infinite-range  Mexican-hat-type  interaction.
We derived two auxiliary equations 
  by expressing the order parameters by the number density of the  oscillators.  
 We used them to   analytically determine the phases of
 the order parameters, 
  derive  self-consistent equations, 
and obtain three non-trivial solutions
that are characterized by the order parameters and the rotation numbers of the synchronized
oscillators $\vecX^* _\theta$s.
We drew  phase diagrams by using formulas for the phase boundaries
derived using the unstable Pn solution, found that the unstable Pn solution differs from
the  paired solution of the stable synchronized solution, 
and the transition due to  pair annihilation of the solution and the 
 relevant solution is a new type of instability that does not
 exist in the Kuramoto model.
We also  analytically obtained the location-dependent
 distribution of the resultant frequencies and entrained phases
 and validated the theoretical results by simulation, 
except for the chaotic behavior of the desynchronized oscillators.
Our numerical results 
 suggest that  the system behaves quasi-periodically in the limit of $N$ infinity. 
\par
In general, when nonlinear dynamical systems that have 
stable limit cycle oscillators are weakly coupled,  the whole system can be
described by phases of oscillators,  and the dynamical equation
 is reduced to the evolution equation 
for phases with general interaction $J_{ij}$.
Therefore, by applying the present method to weakly coupled dynamical systems on a circle,
we should be able to obtain  new types of solutions and new types
 of synchronization--desynchronization phase
transitions.  Several such studies are now underway.
%


\end{document}